\documentclass{aa}
\usepackage[varg]{txfonts}
\usepackage{graphicx}
\usepackage{natbib}
\usepackage{amssymb}
\usepackage{amsmath}
\usepackage{grffile}
\usepackage{longtable,lscape}
\usepackage{color}
\usepackage{pdfpages}
\begin{document}

\title{Shear nulling after PSF Gaussianisation: Moment-based weak lensing measurements with subpercent noise bias}
\titlerunning{SNAPG: Shear nulling after PSF Gaussianisation}
  
\author{Ricardo Herbonnet\inst{1}\thanks{Email: herbonnet@strw.leidenuniv.nl}
  \and Axel Buddendiek\inst{2} 
  \and Konrad Kuijken\inst{1}}

\institute{ Leiden Observatory, Leiden University, PO Box 9513, 2300 RA, Leiden, The Netherlands
  \and Argelander-Institut f\"{u}r Astronomie, Rheinische Friedrich-Wilhelms-Universit\"at Bonn, Auf dem H\"{u}gel 71, 53121 Bonn, Germany }

\date{Submitted \today}

\def\LaTeX{L\kern-.36em\raise.3ex\hbox{a}\kern-.15em
    T\kern-.1667em\lower.7ex\hbox{E}\kern-.125emX}

\newcommand{\x}{{\mathbf x}}
\newcommand{\y}{{\mathbf y}}
\newcommand{\z}{{\mathbf z}}
\newcommand{\A}{{\mathbf A}}
\newcommand{\C}{{\mathbf c}}
\newcommand{\K}{{\mathbf K}}
\newcommand{\G}{{\mathbf G}}
\newcommand{\pmat}{{\mathbf P}}
\newcommand{\bmat}{{\mathbf B}}
\newcommand{\cmat}{{\mathbf C}}
\newcommand{\fmat}{{\mathbf F}}
\newcommand{\hmat}{{\mathbf H}}
\newcommand{\nmat}{{\mathbf n}}
\newcommand{\V}{{\mathbf V}}

\abstract{Current optical imaging surveys for cosmology  cover large areas of sky. Exploiting the statistical power of these surveys for weak lensing measurements requires shape measurement methods with subpercent systematic errors.}{We introduce a new weak lensing shear measurement algorithm, shear nulling after PSF Gaussianisation (SNAPG), designed to avoid the noise biases that affect most other methods.} 
{SNAPG operates on images that have been convolved with a kernel that renders the point spread function (PSF) a circular Gaussian, and uses weighted second moments of the sources. The response of such second moments to a shear of the pre-seeing galaxy image can be predicted analytically, allowing us to construct a shear nulling scheme that finds the shear parameters for which the observed galaxies are consistent with an unsheared, isotropically oriented population of sources. The inverse of this nulling shear is then an estimate of the gravitational lensing shear.} 
{We identify the uncertainty of the estimated  centre of each galaxy as the source of noise bias, and incorporate an approximate estimate of the centroid covariance into the scheme.
We test the method on extensive suites of simulated galaxies of increasing complexity, and find that it is capable of shear measurements with multiplicative bias below 0.5 percent.}{}


\maketitle

\section{Introduction}
The effect that masses can act as lenses and bend the path of light rays is called gravitational lensing. In the weak lensing regime first considered by \citet{firstwl} we statistically measure the slight distortion of the shapes of background galaxies by foreground lenses, called the shear. The subtle effects of weak gravitational lensing on galaxy shapes are an immensely powerful tool in observational astronomy. 
Amongst other applications, weak lensing has been an invaluable tool for cosmology through measurements of shear-shear correlations, called cosmic shear, which are connected to the dark matter power spectrum. 
After its first detection 15 years ago \citep{bacon,vanwaerbeke,wittman00,kaiser_cosmicshear} cosmic shear has been extensively used in cosmological studies  (e.g. \citealt{kilbinger,kids450,des_jarvis}). 

Currently, large ($>$1000 deg$^2$) cosmic shear surveys are ongoing, such as the Kilo Degree Survey \citep{jelte_kids}, the Dark Energy Survey \citep{des}, and Hyper Suprime-Cam \citep{hsc};  more hemisphere-sized missions are planned, such as LSST \citep{lsst}, \textit{WFIRST} \citep{wfirst}, and \textit{Euclid} \citep{euclid}. These surveys will observe unprecedented numbers of galaxies, pushing down statistical errors, and hence requiring percent (for ongoing missions) to subpercent level accuracies (for future missions)  on the measured galaxy shapes.

In order to conduct weak lensing studies a crucial point is to measure the shapes of faint background galaxies with high accuracy as well as high precision in the face of inevitable noise, finite image resolution, and pixel effects.
The first weak lensing techniques used the moments of the galaxy's image to estimate its shape and are known as moment-based methods (e.g. \citealt{kaiser}; \citealt{ksb} (hereafter KSB); \citealt{rhodes00}). These techniques need to use a weighting function with which to cut off the moment integrals so that the moments are not dominated by noise. Having to correct for the effect of the weight function and the PSF convolution are the main challenges for this class of techniques. The widely used KSB method employs an approximate deconvolution scheme, which assumes that the PSF is nearly Gaussian. Newer moment-based methods have improved upon the PSF correction \citep{deimos}, and there have been methods that change the PSF to make the measurement more exact (as explained in \citealt{hirataseljak} and used by  \citealt{mandelbaum_regauss,eholics_era,okura_psfsmearing}).

An alternative class of techniques relies on models of galaxies which are convolved with a PSF and then fit to the galaxy image and are hence known as model-fitting methods \citep[e.g.][]{kuijken99,lensfit,im3shape}. These techniques have the benefit of an accurate treatment of the PSF, but in return require realistic models of galaxies. The model of a galaxy is usually a parametric model (e.g. a linear combination of Sersic profiles) and if it does not resemble the intrinsic galaxy, the results can be biased \citep{bernstein_mb,voigtbridle}. A similar class of techniques, known as shapelets methods \citep{bernsteinjarvis,refregierbacon_shapelets,kuijken_shapelets}, use a set of basis functions which can, in theory, model any galaxy morphology by invoking ever higher order functions. However, in practice the order has to be truncated as the higher functions are dominated by noise, again leading to an unrepresentative galaxy model.
In addition, noise in the galaxy image biases all shape measurement methods due to the non-linear dependence of the galaxy's ellipticity (the usual description of its shape) on the surface brightness \citep{refregier_nb,melchior12,viola14}.

In order to quantify these uncertainties and to find ways of calibrating the different techniques, the weak lensing community started shape measurement challenges in which teams  competed by using their  methods to obtain the most unbiased shear estimate. This started with a general census and benchmark tests in the STEP challenges \citep{step1,step2} and continued with  GREAT challenges \citep{great08,great10,great3}, which focused on the understanding of different sources of bias. After the most recent GREAT3 challenge it appears that the development in shape measurement algorithms is slowly reaching the goals set by ongoing cosmic shear surveys.

The recent improvement in accuracy was mainly due to the advanced understanding of noise bias. Several authors have introduced correction schemes into their shape measurement methods which are able remove a large portion of the noise bias \citep{great3}. An alternative route is to avoid biased shear estimators by using estimators with a linear response to the pixel data instead of traditional non-linear variables, such as the ellipticity. Several authors have used the second moments of the galaxy's image brightness to estimate the shear \citep{zhangkomatsu,bernsteinarmstrong14,viola14}. Recently, \citet{bernsteinarmstrong16} have reported that their Bayesian method based on moments is able to reach subpercent accuracy even with low signal-to-noise ($S/N$) galaxies. However, the drawback of any Bayesian analysis is the requirement of accurate priors, for which external deep observations would be required. This requirement also means that there is no shear estimate for single galaxies, as then knowledge of the intrinsic galaxy profile would be needed, but only a shear estimate for an ensemble of galaxies. 

In this paper we propose a novel shape measurement method which may help to reach the ambitious goals of future cosmic shear experiments. Shear nulling after PSF Gaussianisation (SNAPG) is a moment-based method based on a circular Gaussian PSF and weight function, and requires the images to be preprocessed with a PSF Gaussianisation routine. For such galaxies we have an analytic relation between the moments of the galaxy and the shear.
Shearing a population of galaxies introduces anisotropy to their ellipticity distribution. Using the analytic expressions, SNAPG reintroduces isotropy to this population by applying a nulling shear to the weighted second moments. The inverse of the nulling shear is then the shear estimate. Such a nulling technique was first advocated by \citet{bernsteinjarvis}.
We propose an analytic correction to mitigate the bias due to centroid errors \citep{bernsteinjarvis}, which is directly computed from the galaxy image. 

SNAPG is similar to the \citet{bernsteinarmstrong16} method, but instead of a Bayesian framework it uses a nulling technique extract the shear from the second moments of a population of galaxies. It does not require a prior on the intrinsic moments of the galaxy population, but instead relies on the more general requirement that galaxy ellipticities are isotropic. Our novel method thus only produces a shear value for an ensemble of galaxies, but has the benefit that no auxiliary data is needed.

In Sect. \ref{sec:theory} we introduce the SNAPG concept and a correction for the bias due to centroid errors. 
Section \ref{sec:sims} describes the image simulations we use to test SNAPG, and in Sect. \ref{sec:tests} and Sect. \ref{sec:great08} we present the results of the test runs. This is followed by a detailed discussion in Sect. \ref{sec:discussion}, and a  summary in Sect. \ref{sec:summary}.

\section{Theory}
\label{sec:theory}

\subsection{Principles of SNAPG}
\label{sec:snapg}

Our novel method combines elements from a number of shear measurement methods. It follows KSB and \citet{lk97} in its use of Gaussian-weighted second moments, and uses a nulling technique to estimate the shear \citep{bernsteinjarvis}. We explain the basics of moment-based methods, such as KSB and \citet{lk97} in Sec. \ref{sec:basics}.

Because the ellipticities used in KSB are non-linear functions of the pixel values, pixel noise makes them biased estimators. In SNAPG we work with the second moments of galaxies instead, which even in the presence of pixel noise are unbiased estimators as long as the pixel noise in the image is unbiased (as has been previously explored by \citealt{zhangkomatsu} and \citealt{viola14}). 
For mathematical tractability we require that the PSF in the images is Gaussian and circular; this allows us to work out analytically how the weighted moments respond to any pre-seeing shear (Sec. \ref{sec:shearmom}). A similar exercise was done by \citet{rhodes00}, but here we do not make any simplifying assumptions and find an expression which is valid for all values of the gravitational shear. Note that we do not try to find the intrinsic unweighted moments as in \citet{lk97}; instead, we are only interested in the response of the weighted moments to a shear, similar to \citet{bernsteinarmstrong16}.

The centroid of a source needs to be chosen before the second moments can be calculated. As this position is determined from noisy data, there is a noise dependent shift in the centroid. We show how to incorporate the uncertainty on the centroid of the galaxy into the shear estimator  in the approximation where this uncertainty is distributed as a bivariate Gaussian in Sect. \ref{sec:bias}.

The response of the weighted second moments to shear can be used to find the inverse shear which counteracts the gravitational lensing shear. Hence, given the true shear, we can use the inverse shear to compute the second moments of the galaxy before it was lensed. As the true gravitational lensing shear is unknown, we cannot use each galaxy as an independent shear estimator. Rather, we use an ensemble of sheared galaxies as a probe of systematic alignments and calculate the {nulling shear} that needs to be applied to this ensemble to render their intrinsic ellipticity distribution isotropic. The nulling shear is then the opposite to the true shear affecting these galaxies (Sec. \ref{sec:nulling}). Our approach differs from the approach of \citet{viola14} and \citet{zhangkomatsu}, who average the numerator and denominator of the ellipticity separately to avoid introducing biases. In SNAPG only the numerator of the ellipticity is used as a measure of the isotropy of the ellipticity distribution, and its response to shear calculated in order to null the signal.

Because real PSFs are not circular Gaussians, SNAPG can only be applied to images that have been convolved with a suitable Gaussianisation kernel (Sec. \ref{sec:psfgauss}). Such a convolution is a linear operation on the pixels, so does not introduce noise bias in the second moments. However, it does correlate the pixel noise, the effect of which can be tracked and corrected for.

\subsection{Lensing basics}
\label{sec:basics}
Here we introduce the basic expressions regarding general shear estimation via the ellipticity of a galaxy as we  refer to them often throughout this section. For a more detailed weak lensing review see \citet{bartelmann}.

A gravitational potential changes the path of light rays moving through it, thereby changing the observed direction of incoming light rays. For extended luminous objects different light rays can be deflected differently and thus we will observe a distorted image of a distant object. To the first order this distortion consists of a stretch (shear) and a magnification (convergence). The deflection angle of light rays from the source depends on the gradient of a suitably defined lensing potential, $\Psi$. The relation between the position of the source $\pmb{\beta}$ and the position of the  observed image $\pmb{\theta}$ is known as the lens equation 
\begin{equation}
 \pmb{\beta} = \pmb{\theta} - \nabla \Psi(\pmb{\theta}) .
\end{equation}
Given that the deflection angles in weak lensing are small, the distortion can be expressed in terms of a Jacobian matrix
\begin{eqnarray}
 \mathbf{A} = \frac{\partial \beta_i}{\partial \theta_j} = \left( \delta_{ij} - \frac{\partial^2 \Psi(\pmb{\theta})}{\partial \theta_i \partial \theta_j} \right) 
  =& \begin{pmatrix} 1-\kappa-\gamma_1 & -\gamma_2 \\ -\gamma_2 & 1-\kappa+\gamma_1 \end{pmatrix} \nonumber \\
  \equiv & \dfrac{1}{1-\kappa} \begin{pmatrix} 1-g_1 & -g_2 \\ -g_2 & 1+g_1 \end{pmatrix} .
\end{eqnarray}
The parameters 
\begin{eqnarray}
\kappa =& \dfrac{1}{2} \left(\dfrac{\partial^2 \Psi(\pmb{\theta})}{\partial \theta_1^2} + \dfrac{\partial^2 \Psi(\pmb{\theta})}{\partial \theta_2^2} \right) , \\ \nonumber
\gamma_1 =& \dfrac{1}{2} \left(\dfrac{\partial^2 \Psi(\pmb{\theta})}{\partial \theta_1^2} - \dfrac{\partial^2 \Psi(\pmb{\theta})}{\partial \theta_2^2} \right), \\ \nonumber
\gamma_2 =& \dfrac{\partial^2 \Psi(\pmb{\theta})}{\partial \theta_1 \partial \theta_2} \nonumber
\end{eqnarray} 
are the gravitational lensing convergence $\kappa$ and the two components of the shear $\gamma_1$, $\gamma_2$, respectively. Without information on the intrinsic size of the lensed source, only the reduced shears $g_1,g_2$ can be measured.

The shear affects a galaxy's polarisation according to
\begin{equation}
 \chi^\mathrm{i} = \frac{\chi - 2g + g^2 \chi^*}{1+|g|^2-2\Re(g\chi^*)} ,
 \label{eq:chishear}
\end{equation}
where $\chi$ and $\chi^i$ are the observed and the intrinsic, unlensed polarisation. As the intrinsic shape of a galaxy cannot be measured and the weak lensing shear is very small, the shear has to be statistically obtained from a large number of galaxies experiencing the same distortion. Assuming that galaxies are randomly oriented, the intrinsic polarisations should average out, $\langle \chi^i \rangle =0$.

Moment-based methods construct the polarisation of an object from the second moments of image brightness $Q_{ij}$ 
\begin{equation}
 \chi = \frac{Q_{11}-Q_{22}+2i Q_{12}}{Q_{11}+Q_{22}} .
 \label{eq:polarisation}
\end{equation}

These moments are defined as the noiseless unweighted moments on the intrinsic galaxy image $I^\mathrm{i}(\x)$,
\begin{equation}
Q_{ij}^\mathrm{i} =  \int d\x \, I^\mathrm{i}(\x) x_i x_j W(\x),
\end{equation}
with the weight function $W(\x)=1$.
However, in practice a galaxy is observed convolved with a PSF $P(\x)$,
\begin{equation}
I^\mathrm{o}(\x)=\int d\x' \, I^\mathrm{i}(\x')P(\x-\x'),
\end{equation}
and the weight function $W(\x)$ that goes to zero at large $\x$ is required for the moments not to be dominated by the noise on the image. The aim of moment-based methods is then to estimate the intrinsic polarisations by correcting for the weight function and PSF.

\subsection{Effect of pre-seeing shear on observed Gauss-weighted moments}
\label{sec:shearmom}

In the case when the PSF is Gaussian, we can reconstruct what the second moments would have been if the galaxy had been sheared.

The weighted second moments of the observed image are
\begin{align}
\label{eq:defQ}
Q_{ij}^\mathrm{o} = & \int d\x \, I^\mathrm{o}(\x) x_i x_j W(\x) \\ \nonumber
         = & \int\int d\x \, d\x' \,I^\mathrm{i}(\x')P(\x-\x')x_i x_j W(\x),
\end{align}
with $W(\x)$ a weight function that depends only on $|\x|$.
The order of integration can be swapped and Eq.~\ref{eq:defQ} rewritten as 
\begin{equation}
Q_{ij}^\mathrm{o}=\int d\x'\, I^\mathrm{i}(\x') \left[\int d\x\, P(\x-\x')x_i x_j W(\x)\right]
,\end{equation}
and so relate the weighted second moments directly to the intrinsic galaxy shape as an integral 
weighted by the expression in square brackets (which depends only on the weight function and PSF).

A gravitationally lensed source has a distorted image: the intrinsic image $I^\mathrm{i}(\x)$ is transformed to
\begin{equation}
I^\A(\x)=I^\mathrm{i}(\A\x)
\label{eq:IA}
\end{equation}
by the distortion matrix $\A$.

In order to measure the gravitational shear, we need to know the weighted second moments that we would 
observe if the galaxy had been distorted by a distortion matrix $\A$ before PSF convolution;  these can be written as
\begin{eqnarray}
Q_{ij}^\A & = & \int d\x''\, I^\A(\x'') \left[\int d\x\, P(\x-\x'')x_i x_j W(\x)\right] 
\label{eq:QA1}\\ \nonumber
          & = & \int {\frac{d\x'}{|\det\A|}} I^\mathrm{i}(\x')
\label{eq:QA2} \\ \nonumber
          & & \times \left[\int d\x\, P(\x-\A^{-1}\x')x_i x_j W(\x)\right]
\end{eqnarray}
by means of Eq.~\ref{eq:IA} and the transformation $\x'=\A\x''$.
We now show how the moments $Q^\A_{ij}$ of the sheared source can be derived from the observed, PSF-convolved 
image $I^\mathrm{o}(\x)$, by constructing a new weight function $W^\A_{ij}(\x)$ which satisfies, for arbitrary $\x'$,
\begin{align}
\label{eq:WA}
\int {d\x\over|\det\A|} P(\x-\A^{-1}\x')x_i x_j  & W(\x) \\ \nonumber 
                                                 & = \int d\x P(\x-\x') W^\A_{ij}(\x) .
\end{align}
It is easy to see from Eq.~\ref{eq:QA2} that integrating the observed (PSF-convolved but unsheared) image times 
the weight function $W^\A_{ij}$ will give the moments $Q^\A_{ij}$.  Equation~\ref{eq:WA} shows that $W^\A_{ij}$ can 
be constructed from the original weight function $W$ and the PSF $P$ by the following sequence of operations:
\begin{enumerate}
\item Convolving $x_ix_jW$ with the PSF;
\item Distorting the result of the previous step with distortion matrix $\A$ and divide by $|\det\A|$;
\item Deconvolving the result of the previous step by the PSF.
\end{enumerate}
This recipe is valid as long as the deconvolution in the final step is well defined.

We do not attempt to solve the general problem, but concentrate on the simpler case where both the PSF $P$ and 
the weight function $W$ are round Gaussians:
\begin{equation}
P(\x)={1\over2\pi p^2}e^{-|\x|^2/2p^2}
\end{equation}
and
\begin{equation}
W(\x)=e^{-|\x|^2/2w^2}.
\end{equation}
In Appendix~\ref{app:gconv} we derive an expression (Eq.~\ref{eq:xyggconv}) for the convolution of $G_1$ 
with $x_ix_jG_2$, where $G_1$ and $G_2$ are Gaussians of arbitrary covariance matrices $\pmat$ and $\V$ 
respectively. By substituting $\V=w^2{\mathbf 1}$, $\pmat=p^2{\mathbf 1}$, and $\y=\A^{-1}\x'$ in Eq.~\ref{eq:xyggconv}, we can write the left-hand side of Eq.~\ref{eq:WA} as
\begin{align}
\label{eq:PA}
& \int {d\x\over|\det\A|} P(\x-\A^{-1}\x')x_ix_jW(\x)  \\
= & {w^4\over (w^2+p^2)^2}{e^{-\frac12|\A^{-1}\x'|^2/(w^2+p^2)}\over|\det\A|} \nonumber \\ \nonumber
& \times \left[p^2\delta_{ij}+{w^2\over w^2+p^2}\left(\A^{-1}\x'\right)_i \left(\A^{-1}\x'\right)_j
\right]. \nonumber \qquad \nonumber
\end{align}
The final step is now to deconvolve this expression by the PSF in order to obtain an expression for the
weight function $W^\A_{ij}$ that satisfies Eq.~\ref{eq:WA}. We first calculate the result of convolving 
with a general Gaussian PSF of covariance matrix $\pmat$; the deconvolution we seek is then obtained by 
setting $\pmat=-p^2{\mathbf 1}$ (note the sign).
The first term (involving $\delta_{ij}$) is straightforward: convolving two Gaussians results in a new Gaussian with covariance matrix equal to the sum. The second term can be calculated using the result of Eq.~\ref{eq:xyggconv} in Appendix~\ref{app:gconv} by setting the matrix $\V$ defined there to $\V=(w^2+p^2)\A^2$. 

After some work we find
\begin{align}
\label{eq:waij}
W^\A_{ij}(\x) & = w^4 {e^{-\frac12 \x^T\bmat^{-1}\x} \over |\det \bmat|^\frac12} \\ \nonumber
& \times \left[\delta_{ij} + w^2 \left(
\left(\A\bmat^{-1}\x\right)_i \left(\A\bmat^{-1}\x\right)_j
-\left(\A\bmat^{-1}\A\right)_{ij}
\right)
\right] 
\end{align}
with
\begin{equation}
\bmat=(w^2+p^2)\A^2-p^2{\mathbf 1}.
\label{eq:defB}
\end{equation} 

\begin{figure*}
 \centering
 \includegraphics[width=9.cm,height=9.cm,keepaspectratio=true, trim={2cm 0.5cm 2cm 0,5cm}]{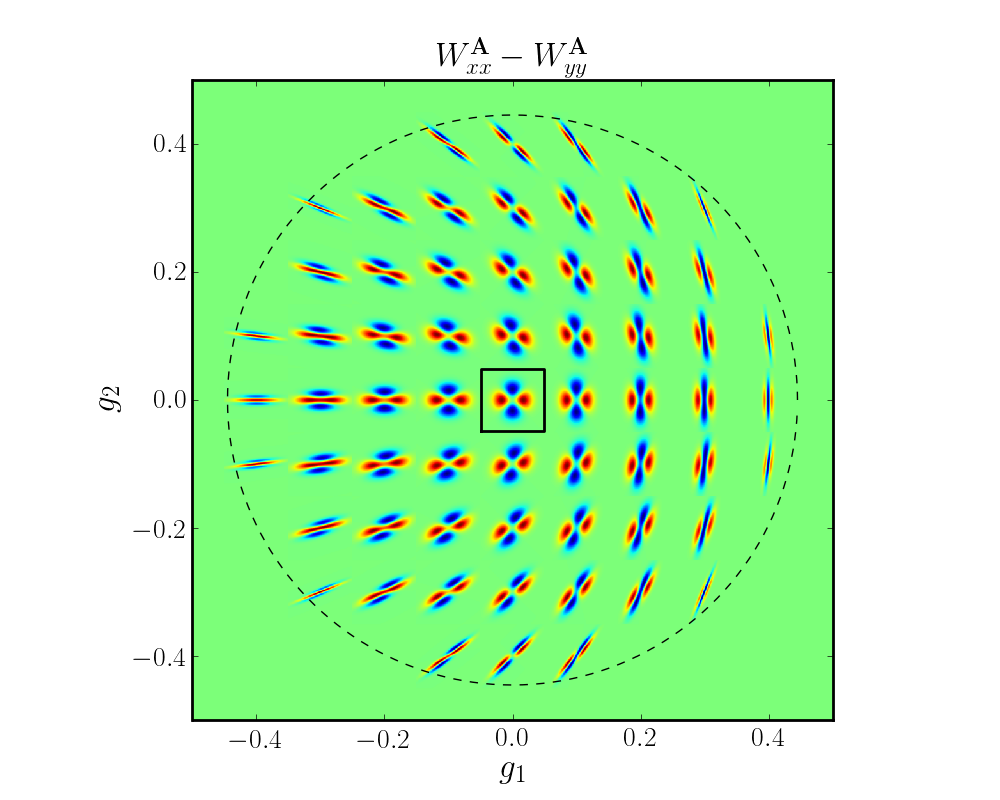}
 \includegraphics[width=9.cm,height=9.cm,keepaspectratio=true, trim={2cm 0.5cm 2cm 0.5cm}]{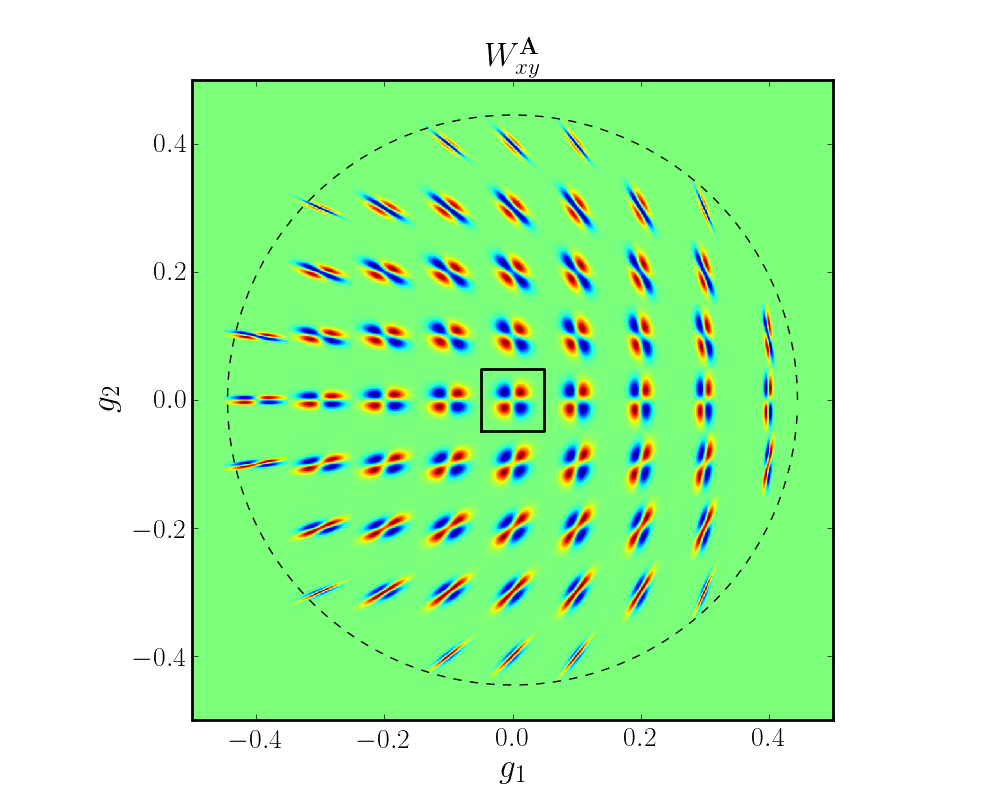}
 \caption{ Example shear filter functions $W_{ij}^{\A}$, for filter and PSF size $w=3$, $p=2$, ($\kappa=0$). Integrating an image multiplied with the filter function for a particular shear $(g_1,g_2)$ yields the weighted second moments the image would have had if it had been sheared by that amount before seeing convolution. The central box is 25 units on a side, and the dashed line indicates the maximum shear value that can be applied for the given $w$ and $p$ (see Eq.~\ref{eq:wfsize}). Red is positive, blue is negative. 
Left: Filters corresponding to $Q_{xx}-Q_{yy}$. Right: Filters corresponding to $Q_{xy}$.}
 \label{fig:weightfunction}
\end{figure*}

The weight function $W^\A_{ij}$ is only useful in practice if it tends to zero at large $|\x|$. This is 
the case as long as the distortions are small enough so that both eigenvalues of $\bmat$ are positive, 
which is true when 
\begin{equation}
\kappa+\gamma < 1-{p\over\sqrt{w^2+p^2}} .
\label{eq:wfsize}
\end{equation}
As long as the weight function is wider than the PSF ($w>p$, a reasonable choice if one wants to avoid 
unnecessarily noisy measurements), this means a useful $W^\A_{ij}$ can be constructed for $\kappa+\gamma$ 
at least up to 0.3.
We show the form of the weight function $W_{ij}^{\A}$ for a grid of $(g_1, g_2, \kappa=0)$ in Fig. \ref{fig:weightfunction}.

\subsection{Bias as a consequence of centroiding errors}
\label{sec:bias}

Applying the filter $W^\A_{ij}$ in Eq.~\ref{eq:waij} to an image yields unbiased estimates for the 
post-seeing weighted second moments of the source about $\x={\mathbf 0}$ as long as the noise on each 
pixel is unbiased. However, in reality the true centre of the source is unknown, and must be estimated 
from the image itself. The associated scatter in the centroid biases the second moments. In this 
section we quantify that bias.

Suppose that $\x_c$ is a noisy estimate of the centroid of the observed image $I^o$. Then, using this 
centroid our estimate for $Q_{ij}^\A$ is
\begin{equation}
\tilde Q_{ij}^\A = \int d\x \, I^\mathrm{o}(\x) W^\A_{ij}(\x-\x_c).
\end{equation}
If the error distribution of the centroids is $f(\x_c)$ then the expectation value of $\tilde Q^\A_{ij}$ is
\begin{align}
\left\langle\tilde Q_{ij}^\A\right\rangle = & \int d\x_c f(\x_c) \tilde Q_{ij}^\A \\ \nonumber
                                          = & \int d\x \, I^\mathrm{o}(\x) \int d\x_c f(\x_c) W^\A_{ij}(\x-\x_c),
\end{align}
i.e. the weight function that determines $\tilde Q^\A_{ij}$ is the original weight function $W^\A_{ij}$ 
convolved with the centroid error distribution $f$. Hence, conversely, an unbiased estimate of $Q^\A_{ij}$ 
is obtained by using a weight function $\hat W_{ij}^\A$ obtained by {de}convolving $W^\A_{ij}$ by the 
centroid error distribution $f$. We assume that $f$ is Gaussian, of covariance $\cmat$. Remembering that expression \ref{eq:waij} for $W_{ij}^\A$ was 
itself obtained by deconvolving Eq.~\ref{eq:PA} by the PSF, we see that $\hat W^\A_{ij}$ is the 
deconvolution of Eq.~\ref{eq:PA} by $P\otimes f$, i.e. by a Gaussian of covariance matrix 
$p^2{\mathbf 1}+\cmat$. As noted under Eq.~\ref{eq:defB}, this deconvolution is simply accomplished by 
using Eq.~\ref{eq:waij} with a modified $\bmat$ matrix
\begin{equation}
\hat \bmat = (w^2+p^2)\A^2-p^2{\mathbf 1} -\cmat.
\label{eq:bhat}
\end{equation}

It remains for us to quantify the covariance matrix $\cmat$ of the centroid error for a given source. 
This will depend on the recipe used to determine the centre.

We centre each source by finding the peak of the correlation of its (noisy) image $I^\mathrm{n}(\x)$ with a 
suitable centring kernel $f$, equivalent to finding the optimum positional match between $I^\mathrm{n}$ and $f$. 
The centroid $\C$ found this way satisfies
\begin{align}
0 = & {\partial\over\partial c_i} \int I^\mathrm{n}(\x)f(\x-\C)d\x \\ \nonumber
  = & \int I^\mathrm{n}(\x)(f_{,i}(\x)-c_jf_{,ij}(\x)+...)d\x
\end{align}
where we used a Taylor expansion on $f$ about $\C={\mathbf 0}$, assumed to be the true centre of our source. 
To derive the noise properties of $\C$ we first separate $I^n(\x)$ into the noise-free observed image 
$I^o(\x)$ and a noise field $\Delta(\x)$, and obtain the first-order relation between $\Delta$ and $\C$:
\begin{equation}
c_j \int I^\mathrm{o}(\x)f_{,ij}d\x = \int \Delta(\x)f_{,i}(\x)d\x .
\label{eq:ctrnoise}
\end{equation}
To calculate the covariance $\overline{c_kc_l}$ we define $F_{ij}$ as the integral on the left-hand 
side of Eq.~\ref{eq:ctrnoise}, and we
assume the background-limited case in which pixel noise is stationary, of constant covariance matrix 
$N(\x-\x')$ across a source image.
Squaring Eq.~\ref{eq:ctrnoise} and averaging over all possible noise realisations then yields
\begin{equation}
\label{eq:ctrcov}
\overline{c_kc_l} F_{ik}F_{jl} =\int\int f_{,i}(\x)f_{,j}(\x')N(\x-\x')d\x d\x' \equiv H_{ij}, 
\end{equation}
and hence the covariance matrix $\cmat$ needed in Eq.~\ref{eq:bhat} is given by
\begin{equation}
\label{eq:cmat}
\cmat=\fmat^{-1}\hmat\fmat^{-1} .
\end{equation}
Here, $\hmat$ depends only on the kernel function $f$ and the pixel noise properties $N$, 
and $\fmat$ can be estimated from the noisy image $I^n$.
If the kernel $f$ is circular and the noise covariance matrix isotropic $N(\x-\x')=\sigma_n^2 \delta (\x -\x')$, where $\sigma_n$ is the root mean square of the noise background, then $\hmat$ becomes a scalar.

A convenient choice is a Gaussian 
\begin{equation}
f(\x)=e^{-|\x|^2/2 a^2}
\end{equation}
for which
\begin{equation}
H_{ij}={\pi \sigma^2_n \delta_{ij} \over 2} 
\end{equation}
and
\begin{equation}
f_{,ij}(\x) = \left( \frac{x_i x_j}{a^4} - \frac{\delta_{ij}}{a^2} \right) e^{-|\x|^2/2 a^2}.
\end{equation}

\subsection{ SNAPG shear nulling estimator}
\label{sec:nulling}

In the previous sections we  constructed the filter $W^\A_{ij}$ which, when applied to an observed Gaussian-PSF smeared image, yields the Gauss-weighted second moments $Q^\A_{ij}$ that would have been observed (with the same PSF) had the galaxy been distorted by distortion matrix $\A$. We have also quantified the noise bias on $Q^\A_{ij}$ due to centroiding errors, and constructed a modified filter $\hat W^\A_{ij}$ that compensates for it. In what follows we will drop the ``hat'' notation and assume that the centroid error correction is applied.

The weight function $W^\A_{ij}$ can be used to construct a shear estimator. If a galaxy is sheared by some known distortion matrix $\A$, then we can use the inverse of $\A$ to find the intrinsic second moments of the galaxy. For a large ensemble of galaxies their combined intrinsic ellipticities (or equivalently their Stokes parameters) average out to zero. Then the search is for the distortion matrix $\A$ which can null the Stokes parameters $(Q^\A_{11}-Q^\A_{22}, 2Q^\A_{12})$ of a sheared population of galaxies. 
The inverse of that distortion matrix is a good estimator of the shear those galaxies experience.
To efficiently search for the distortion matrix we use a nulling scheme similar to one already  used  in shape measurements \citep{bernsteinjarvis}. 
In practice, a trial distortion matrix $\A$ is chosen and the corresponding weight function $W_{ij}^\A$ is computed (see Fig. \ref{fig:weightfunction}),  with which the Stokes parameters for the ensemble of galaxies are calculated. Based on the (an)isotropy of the Stokes parameters a new distortion matrix is chosen, and the previous steps are repeated to reassess the isotropy. This procedure converges in roughly four trials, after which the inverse of the distortion matrix is taken as the shear estimate.

Because galaxies have a wide range of brightness, the Stokes parameters of a galaxy population have a large variance. This translates into a large variance in the nulling shear and increasing precision would require large numbers of galaxies. Alternatively, the moments could be weighted by flux or size to reduce the variance, but this would introduce a bias in the shear. In our current tests such a weight is not required, but we discuss possible solutions for future work in Sec. \ref{sec:precision}.

\subsection{PSF Gaussianisation}
\label{sec:psfgauss}

As indicated in the beginning of Sec.~\ref{sec:shearmom}, SNAPG relies on the assumption of a circular Gaussian PSF. Such a PSF is never present in observational data and thus we need to transform the actual observed PSF into the required PSF. We employ a Gaussianisation process which creates a circular Gaussian PSF by convolving the observed PSF with an appropriate kernel.

Gaussianisation starts by creating a shapelets model of the PSF. Shapelets are a set of basis functions of Gauss-Hermite polynomials, which can be linearly combined to model astronomical objects \citep{refregier_shapelets}. Convolution in shapelet space is a straightforward procedure, making shapelets an ideal basis for the Gaussianisation process. We use the shapelet implementation of \citet{kuijken_shapelets} to create a shapelet model of the PSF. In practice, bright stars can be used to obtain a model of the PSF.
A best fit circular Gaussian of the shapelets model of the PSF is determined. Then a convolution kernel is found that convolves the PSF into the best fit Gaussian. The resulting kernel is applied to the whole image to create galaxies with circular Gaussian PSFs. See \citet{kids15} for more detail on the process of PSF Gaussianisation.

It is worth noting that this procedure is different from the one presented by \citet{hirataseljak}.  They assume a Gaussian form for the intrinsic shape of the galaxy when calculating the corrections for PSF non-Gaussianity, whereas our procedure is valid for any galaxy morphology. However, it does rely on well-sampled data and was designed with only ground-based PSFs in mind. It is unclear how the procedure would perform for diffraction-limited space telescopes.

The convolution mixes information from neighbouring pixels and hence introduces a correlation between the noise on different pixel values. The resulting noise covariance matrix $N(x-x')$ is given by the original image's pixel variance, multiplied by the auto-correlation function of the convolution kernel. It is important to propagate this noise covariance into the centroid error estimate (Eq.~\ref{eq:ctrcov}).

\section{Image simulations}
\label{sec:sims}

\begin{table*}
 \caption{Overview and specifications of all simulated images used to test the performance of SNAPG.}
 \label{tab:sims}
 \begin{tabular}{clccccccc}
 & Set & & PSF & Galaxy type & PSF HLR &  Galaxy HLR & $S/N$ \\ 
  \hline 
  & Well resolved & & Gaussian & Exponential & 1.76 pixels & 2.5 pixels & $\sim$5 - 100 & \\
  & Barely resolved & & Gaussian & Exponential & 1.76 pixels & 1.5 pixels & $\sim$5 - 100 & \\  
  & GREAT08 RNK & & Gaussianised Moffat & Exponential or de Vaucouleurs & 1.72 pixels & 2.1 or 10 pixels & $\sim$200 &  \\
  & GREAT08 LNK & & Gaussianised Moffat & Exponential or de Vaucouleurs & 1.72 pixels & 2.1 or 10 pixels & $\sim$20 & \\
  \hline
 \end{tabular}
\end{table*}

To test the performance of SNAPG we create simulated images of galaxies with known applied shear. Following the image simulations of the GREAT challenges, we create a grid of isolated galaxies on postage stamps. This approach  gives us a clean test of the performance of SNAPG without introducing errors related to blended galaxy isophotes (see \citealt{hoekstra_clusters} for a discussion on how blends affect shear measurements). The images of the GREAT challenges do not have circular Gaussian PSFs, so for a clean test of the SNAPG framework we use \texttt{GalSim} \citep{rowe} to create our own image simulations with perfect circular Gaussian PSFs.

Our simulated galaxy images are a grid of 100 x 100 galaxies, all with exponential profiles of the same size. The grid of postage stamps have a single galaxy randomly offset from the centre of the stamp. The postage stamp is  large enough to avoid any bias due to truncation of the surface brightness profile. The half light radius (HLR) of the galaxies is 2.5 pixels. The flux of all galaxies is the same, so that when noise is added all galaxies will have the same $S/N$. The modulus of the ellipticity of a galaxy is randomly drawn from a Rayleigh distribution of width 0.25, cut off at 0.6 to avoid artificial truncation by the edge of the postage stamp. The position angle of the galaxy is taken from a random uniform distribution between 0 and 180 degrees. The galaxy models are convolved with a Gaussian PSF with a half light radius of 1.76 pixels. The size of the galaxies is larger than the size of the PSF, so we call this set of images the \textit{well resolved} sample. Each image has a constant shear applied to all 10000 galaxies, where the shear is taken from a grid of $(-0.04,-0.035,-0.03,...,0.04)$ for each shear component separately, resulting in 289 different $g_1,g_2$ pairs.

We also create a similar set of images where the galaxy half light radius is set to 1.5 pixels. The half light radius of the Gaussian PSF is 1.76 pixels, so that the PSF is larger than the galaxy. We call this set of images the \textit{barely resolved} sample. Fluxes are fixed and the ellipticity is sampled in the same way as described above. Here too, constant shears are applied to all galaxies on an image, and the shear is taken from the same grid.

These two suites of image simulations contain a total of $5.78 \cdot 10^6$ galaxies. These galaxy images do not contain any noise, instead Gaussian noise is added as required for each test. Each set represents a different target for shape measurement.  The well resolved images present our fiducial dataset as the galaxy shapes are not badly affected by the PSF and provide us with a benchmark test of the performance of SNAPG. 
The barely resolved images present a challenging sample, as galaxy shapes are heavily influenced by pixelisation and severely blurred by the PSF. These galaxies are a difficult target for most shape measurement methods and are sometimes cut from the sample owing to the uncertainty in the galaxy shapes. However, faint small galaxies are abundant in observations and their removal presents a serious loss of statistical power. Having a shear measurement technique able to reliably measure such objects will be a huge advantage for future weak lensing experiments.

\begin{figure}
 \centering
 \includegraphics[width=8.5cm,height=8.5cm,keepaspectratio=true, trim={1cm 0 0 0}]{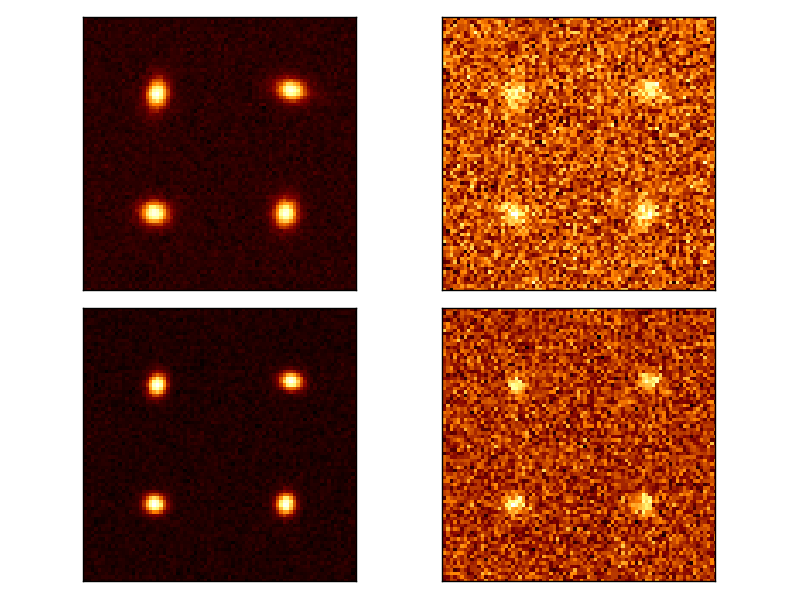}
 \caption{Examples of the simulated galaxies at different noise levels to help visualise varying $S/N$ levels, and well resolved in contrast to barely resolved galaxies. \textit{Top}: Cut-out of the well resolved sample of images for $S/N\approx100$ (left) and $S/N\approx10$ (right). \textit{Bottom}: Cut-out of the barely resolved sample of images for $S/N\approx100$ (left) and $S/N\approx10$ (right). The shape of the barely resolved galaxies is rounded by the PSF and for low $S/N$ both well resolved and barely resolved galaxies the shapes are very much affected by noise.}
 \label{fig:image_sims}
\end{figure}

As a visual aid to interpreting the different sets of simulated galaxy images we show some of our mock galaxy images in Fig. \ref{fig:image_sims}.
The upper images show the fiducial well resolved galaxies and the lower images show galaxies from the challenging set of barely resolved galaxies. The images have Gaussian noise added so that the mean $S/N \approx$ 100 (left panel) and $S/N \approx$ 10 (right panel), where $S/N$ is defined as the \texttt{FLUX/FLUXERR} measured by SExtractor on default settings \citep{sextractor}. A summary of the image properties can be found in Table \ref{tab:sims}.

The images of the GREAT08 challenge \citep{great08} provide us with a test of the PSF Gaussianisation. In addition, we can compare the performance of SNAPG to other tested methods. We use the 15 LowNoise\textunderscore Known (LNK) and 300 RealNoise\textunderscore Known (RNK) sets of images from the challenge, where each image has 10000 isolated galaxies in postage stamps of 40 pixels across. All 10000 galaxies in an image have the same shear applied to provide ample statistics. The galaxies are either an exponential or a de Vaucouleurs profile with a fiducial $S/N=200$ for LNK and $S/N\approx20$ for RNK. The sizes of galaxies are set so that the PSF convolved galaxy size is 1.4 times larger than the PSF size. We use the PSF Gaussianisation algorithm explained in Sec.~\ref{sec:psfgauss} to outfit the GREAT08 images with a circular Gaussian PSF. The Gaussianisation algorithm is applied to PSF \texttt{set0001}, which is a Moffat profile of full width half maximum 2.85 truncated at 5.7 pixels and ellipticity components $e_1=-0.019$ and $e_2=-0.007$. The main properties of the GREAT08 images are summarised in Table \ref{tab:sims}.

\section{Test runs}
\label{sec:tests}
The new shear nulling method is coded in \texttt{python} and we apply it to the image simulations described in the previous section. The code returns the shear value $g_i$ that nulls the average distortion of each 10,000-galaxy image.
First we test the SNAPG formalism, then the centroid bias correction formalism, and finally a full implementation of SNAPG. Throughout this section we use $a$=3 and $w$=3 for the widths of the centroid and moment weight functions, respectively. As noted above, none of the images contains noise. Instead, we add noise for each test as storing each noise realisation presented storage problems. For every level of added noise we calculate the mean signal-to-noise ratio ($S/N$) using SExtractor with default settings and defined as \texttt{FLUX/FLUXERR}. Each measurement we present was obtained by using the full set of 2.89 $\cdot 10^6$ galaxies in each set of simulations described in the previous section. 

The performance of SNAPG is measured by performing a linear fit using the functional form $g_{i,\mathrm{out}} = (1+ m_i )\hspace{1pt} g_{i,\mathrm{true}} + c_i$ \citep{step1} for each shear component $g_i$. This procedure quantifies the shear bias as a multiplicative term $m_i$ (e.g. arising from  method assumptions or noise) and an additive term $c_i$ (arising from imperfect corrections for the elliptical PSF). Because our simulations are ideal with a circular Gaussian PSF we do not expect any additive bias in these tests.

\subsection{High signal-to-noise tests of SNAPG}
\begin{figure*}
 \centering
 \includegraphics[width=12cm,height=9cm,keepaspectratio=true]{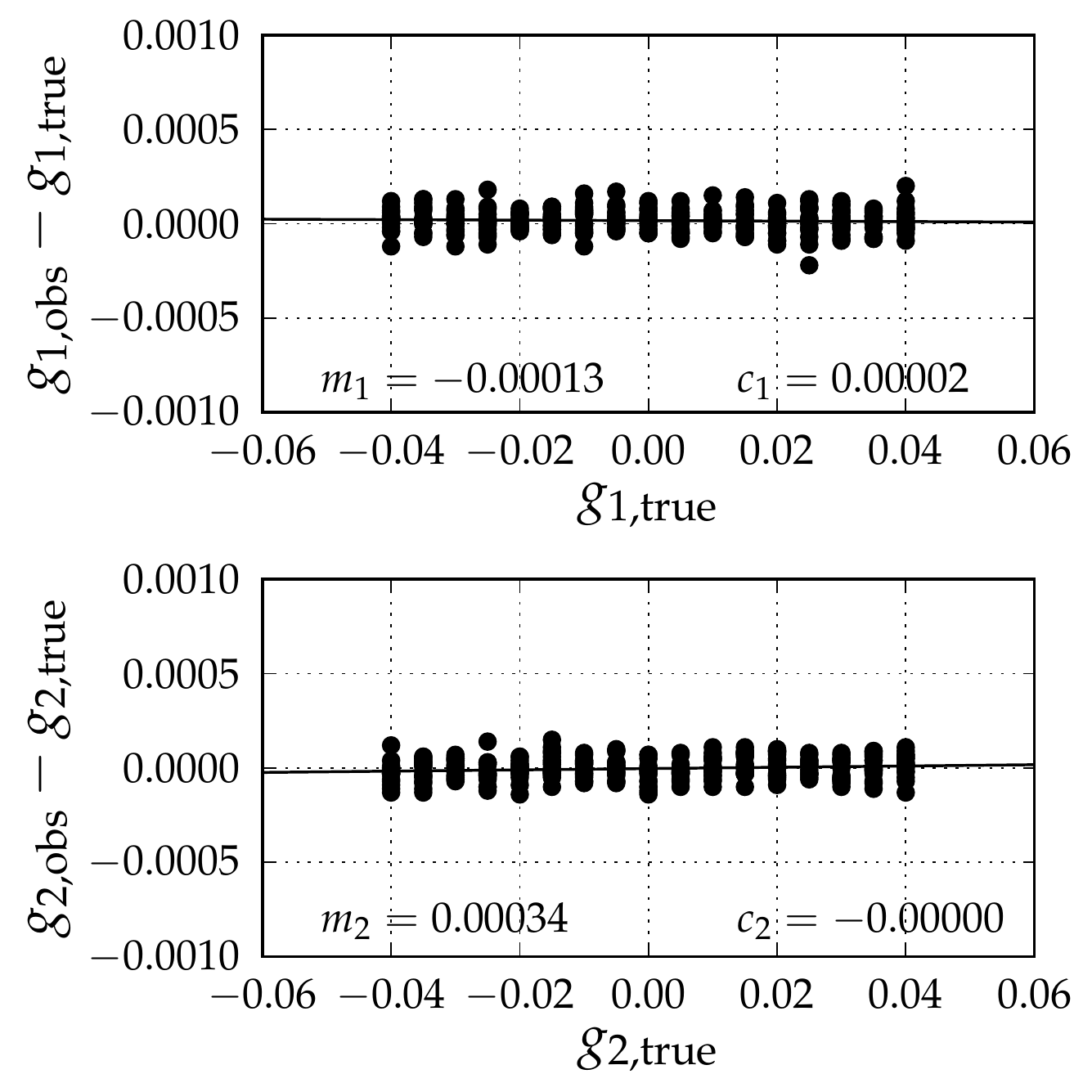}
 \includegraphics[width=12cm,height=9cm,keepaspectratio=true]{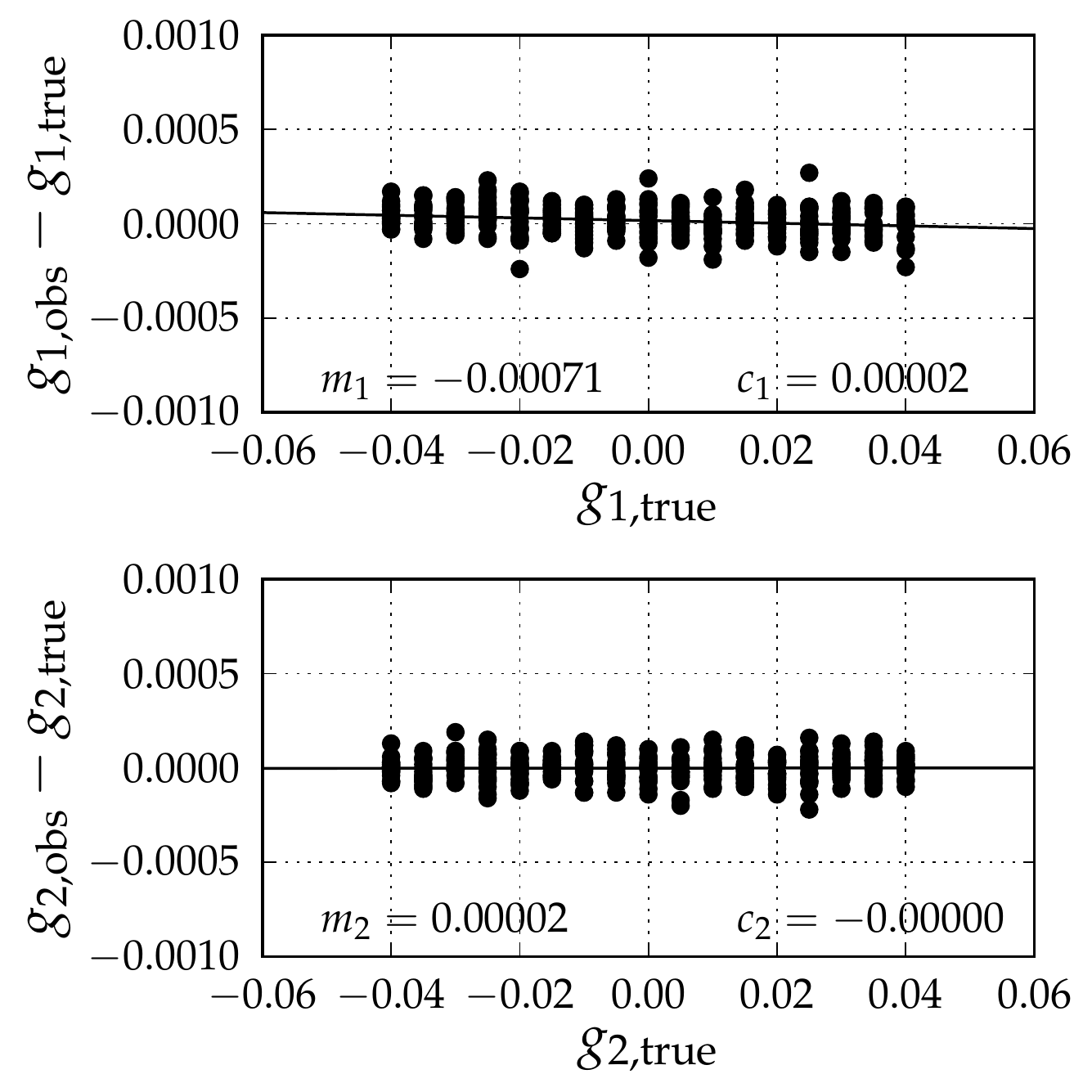}
 \caption{Shear estimates $g_{out}$ measured by SNAPG on images with well resolved (left) and barely resolved (right) galaxies of  $S/N\approx 100$ compared to the input shears $g_{true}$ of the images. Each datapoint is the shear estimated from the 10000 isolated galaxies on an image. The text in the figures shows the multiplicative bias $m$ and the additive bias $c$ of the measurement obtained from the best linear fit shown in black. The potential of SNAPG as a shear measurement method is clear as the true shears can be recovered to an accuracy of less than one part per thousand.}
 \label{fig:high_sn_res}
\end{figure*}

We start by quantifying the performance of SNAPG on the fiducial set of images with well resolved galaxies for $S/N$=100 for the true centroid of the galaxy. In the left panel of Fig. \ref{fig:high_sn_res} we show the measured residuals between the input shear and the measured one,  and the true input versus the measured shear. We find $\langle m \rangle = (+1.6\pm 1.9) \hspace{1pt} 10^{-4}$ and $\langle c \rangle =(-0.1\pm0.1) \hspace{1pt} 10^{-4}$ for the average of the two components of the shear. We also test the algorithm on the set of barely resolved galaxies and the result is plotted in the right panel. Again SNAPG retrieves the applied shear without detectable bias: $\langle m \rangle =(-0.9\pm 2.3) \hspace{1pt} 10^{-4}$ and $\langle c \rangle =(-0.1\pm0.1) \hspace{1pt} 10^{-4}$.

As expected SNAPG returns unbiased shear estimates for our ideal images with circular Gaussian PSF with high $S/N$, showing that the pipeline works. These tests also show the potential of SNAPG as a shear measurement method, regardless of the size of the galaxy in relation to the size of the PSF. We note that at this high $S/N$ the results remain unchanged if the centroid is measured from the data.

\subsection{Tests of centroid bias correction}

We expect a bias in the shear estimate to originate from the random error on the measured centroid due to image noise. In order to test the effectiveness of the centroid bias correction proposed in Sec. \ref{sec:bias}, we perform a test with truly random centroid values. The well resolved images are re-analysed with SNAPG, but the centroids are artificially offset from the true centroid by a random Gaussian value. The error on the centroid is taken from a normal distribution with a standard deviation of 0.5 pixels. Such a distribution would occur for our simulated images with a $S/N \approx 5.5$ for $a=3$.

\begin{figure}
 \centering
 \includegraphics[width=9cm,height=8cm,keepaspectratio=true]{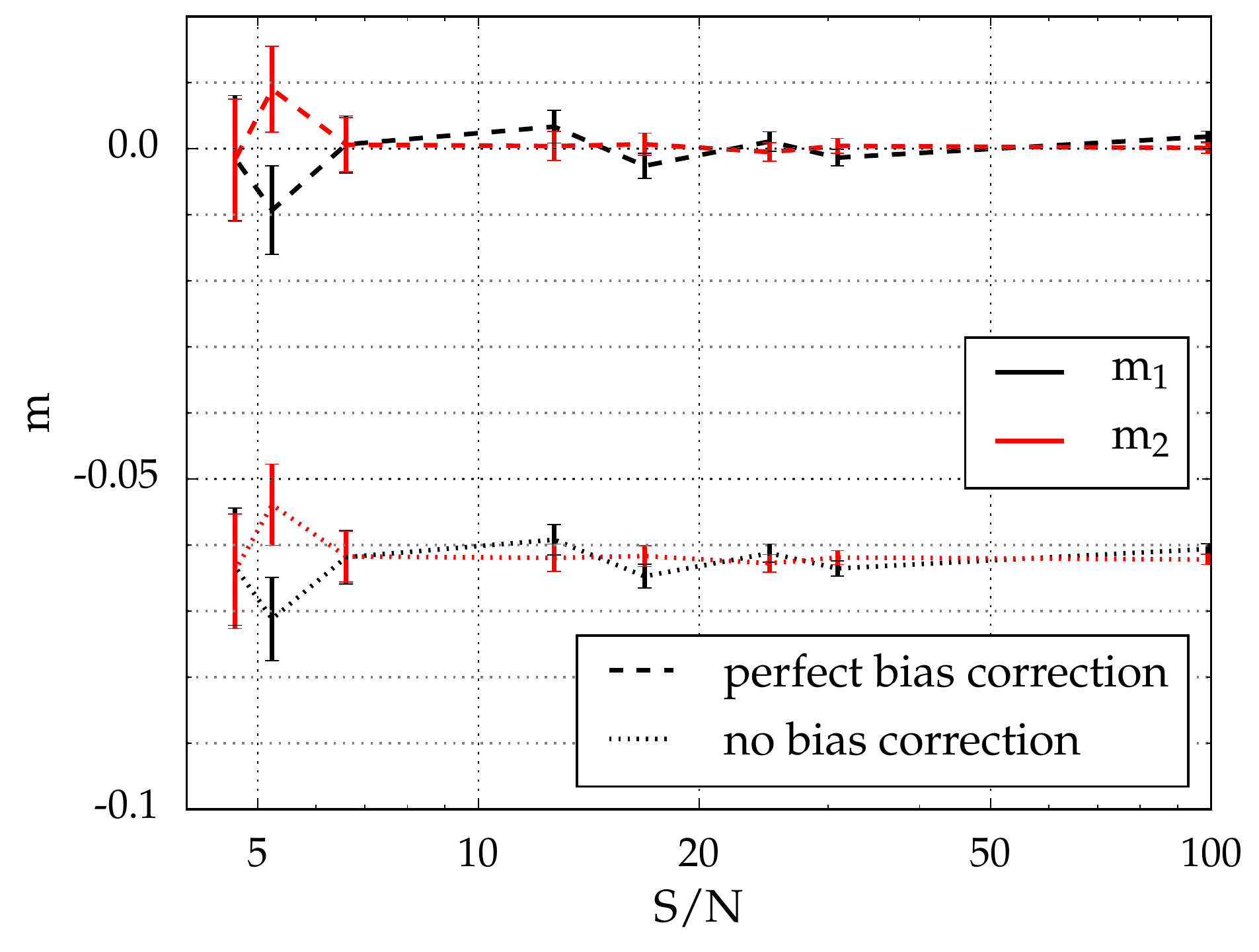}
 \caption{Galaxies in the well resolved image set are assigned random centroid errors, then Gaussian noise is added, after which the shear is measured with SNAPG. Measured multiplicative bias in the shear is plotted versus the mean $S/N$ of all galaxies on the images, with $m_1$ in black and $m_2$ in red. Each datapoint is based on the shear estimates of all 2.89 million galaxies in the well resolved image set. SNAPG results with the exact centroid covariance matrix are shown as dashed lines and results without centroid error correction as dotted lines. The bias from misplaced centroids can be reliably removed by SNAPG, regardless of the noise added to the images.}
 \label{fig:m_sn_xyvar}
\end{figure}

Besides introducing random centroid errors, we also add Gaussian noise to the images before measuring the shear. The addition of noise, which is uncorrelated to the centroid error, should not bias SNAPG as the moments are linear with respect to the noisy surface brightness. Hence, we analyse the images with SNAPG several times where each time Gaussian noise with a different root mean square is added to the images. The results of these tests are presented in Fig. \ref{fig:m_sn_xyvar} as the multiplicative bias $m_1$ in black and $m_2$ in red versus the mean $S/N$ of all galaxies. The dotted lines show the measured bias when no correction is applied. The dashed lines show the bias when the covariance matrix is set to the correct centroid covariance $\cmat=0.5^2 \mathbf{1}$. As we expected, the application of the correction reduces the multiplicative bias from centroid errors to subpercent levels regardless of the noise on the galaxy images. Higher levels of noise increase the variance of the measurements but do not lead to a bias. The mean corrected multiplicative bias over all $S/N$ is $\langle m \rangle = -4.7 \cdot 10^{-4}$ and the measured additive bias is below $10^{-3}$ for all $S/N$.

\subsection{Full test of SNAPG}
\begin{figure}
 \centering
 \includegraphics[width=9cm,height=8cm,keepaspectratio=true]{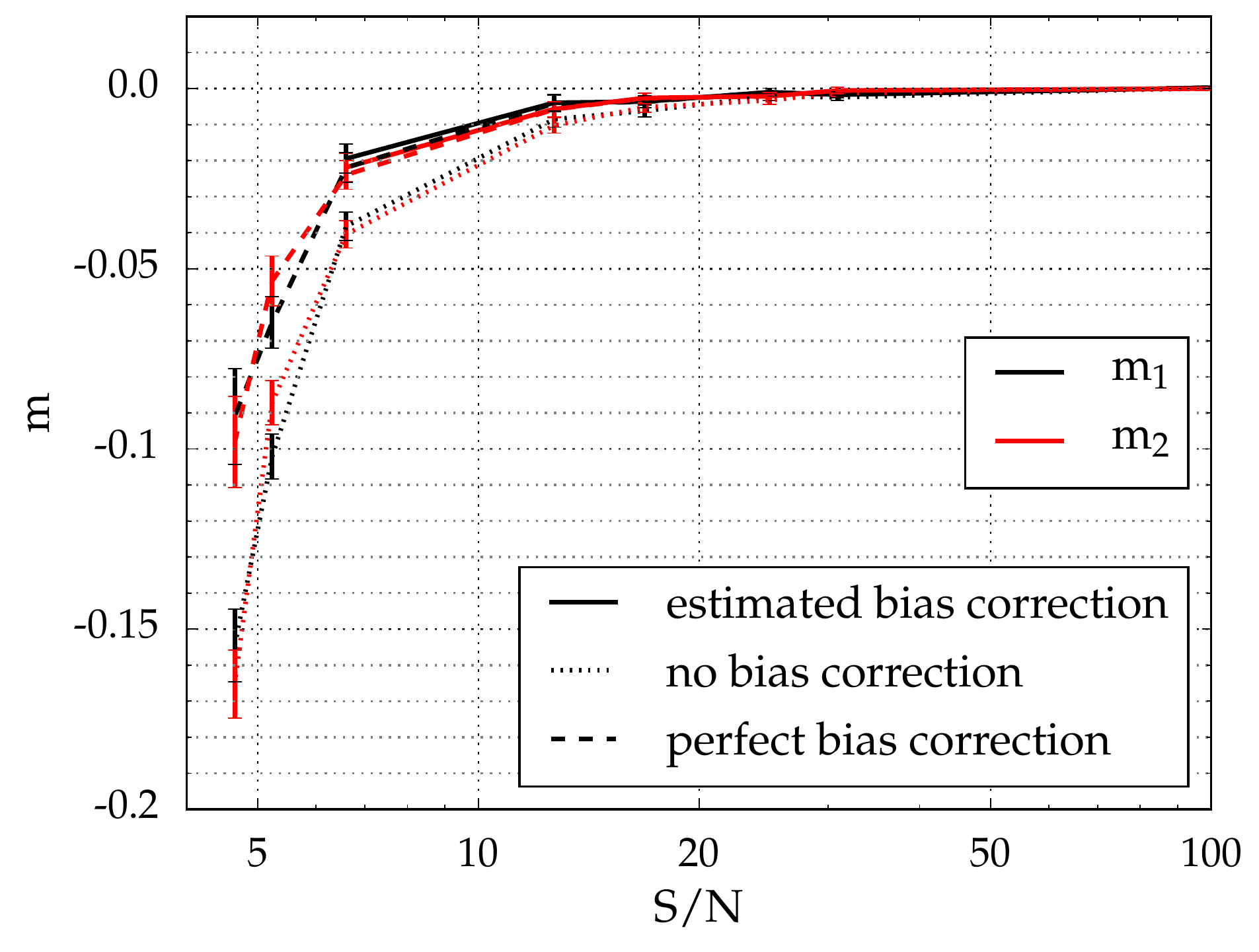}
 \caption{Similar set-up to that in Fig. \ref{fig:m_sn_xyvar}. Dotted lines show the SNAPG results without using the centroid error correction and solid lines show the results for covariance matrices measured from the data. The centroid bias algorithm breaks down for very low $S/N$, so that the solid curves do not reach to $S/N<6$. Dashed lines show the SNAPG results when the mean centroid variance in each image is taken to be the covariance matrix for that image. As the dashed and solid lines are almost indistinguishable, the measured covariance matrix is a good estimate of the true centroid covariance. However, it is unable to remove all bias caused by noise, leaving residual biases of percent level for very faint $S/N<10$ galaxies.}
 \label{fig:m_sn_measctr_res}
\end{figure}

We now add a centroid measurement algorithm to SNAPG and use it as input for SNAPG. The centroids of each galaxy are estimated by nulling the first moments of the galaxy and the centroid error covariance matrix is estimated using Eq.~\ref{eq:cmat}. Gaussian noise is added to the well resolved images and SNAPG is run to obtain a shear estimate. In contrast to the previous test, the noise is now directly related to the centroid error. Again we repeat this exercise for different noise levels and show the multiplicative bias as a function of the measured mean $S/N$ in Fig. \ref{fig:m_sn_measctr_res}. 
The dotted line shows the bias in $g_1$ in black and $g_2$ in red without applying the centroid bias correction and the solid line shows the corrected bias. Centroid errors lead to a bias of several percent for $S/N\approx10$, which is decreased to $\sim1$\% by the centroid bias correction.  The measured additive bias is below 0.1\% for all $S/N$. For very low $S/N$ the estimate for the covariance matrix becomes dominated by noise and the formalism breaks down.

The centroid error correction removes a large part of the noise bias, but does not remove the bias completely. This can occur if the measured covariance matrix is not a good representation of the true centroid variance. To check this hypothesis, we compute the true centroid variance and compare its performance to our previous results. First we measure the centroid from a noisy image and compare this to the true centroid to compute the centroid error $\Delta \x$. We then estimate the centroid variance as the average $C_{ij} = \langle \Delta x_i \Delta x_j \rangle$ over all 10000 galaxies in each image, square the centroid error, and compute the centroid variance as the squared centroid error averaged over all 10000 galaxies in an image $C_{ij} = \langle \Delta x_i \Delta x_j \rangle$. This true centroid error covariance is set into Eq. \ref{eq:bhat} as the covariance matrix for all 10000 galaxies in the image and the shear for the image is measured. We show these results as dashed lines in Fig. \ref{fig:m_sn_measctr_res} and note that they are very similar to our previous results (solid lines). These findings indicate that the estimate of the centroid variance is good, but that there is unresolved noise bias in SNAPG.

\begin{figure}
 \centering
 \includegraphics[width=9cm,height=8cm,keepaspectratio=true]{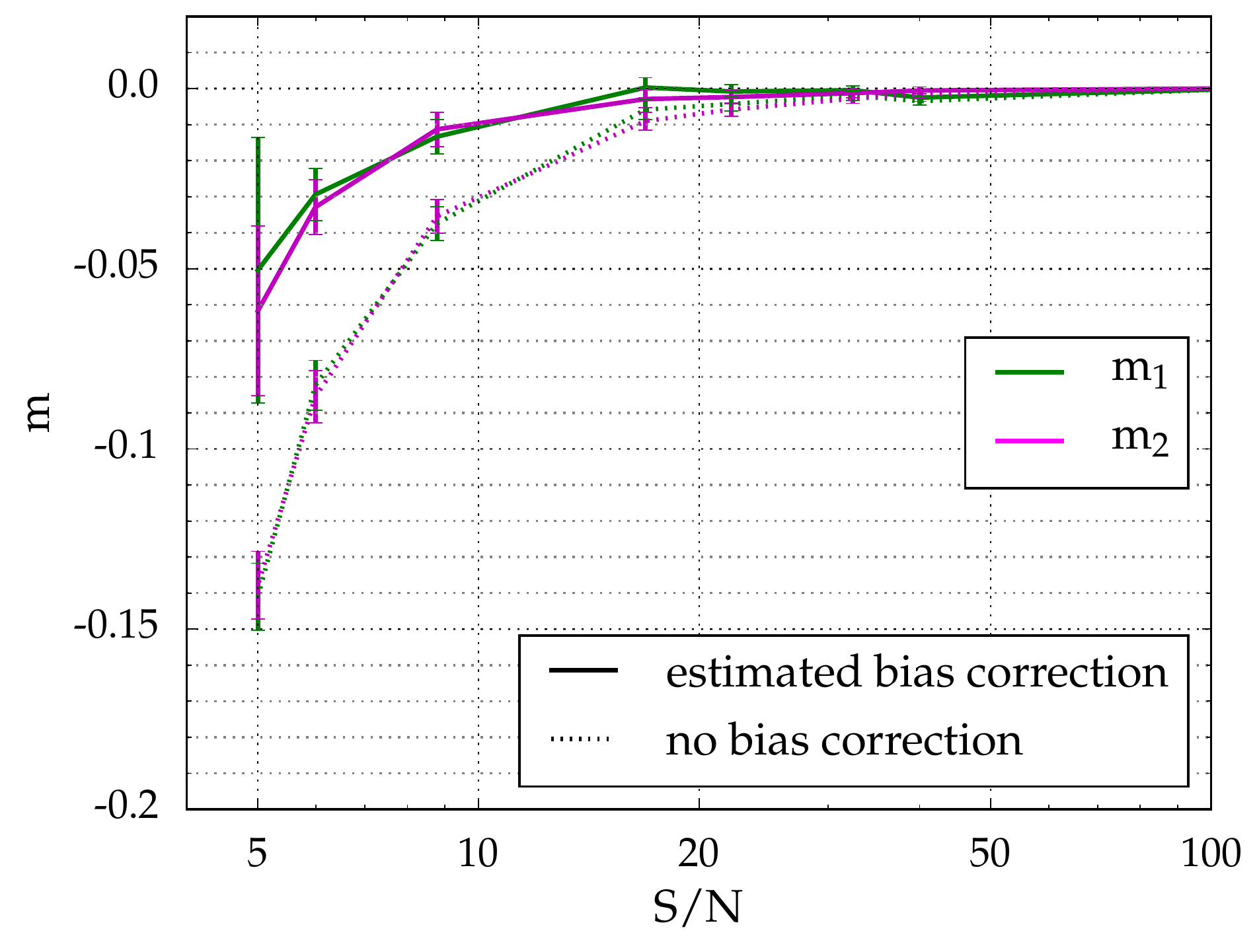}
 \caption{Same as Fig. \ref{fig:m_sn_measctr_res}, but now for the set of barely resolved galaxies. Again SNAPG shows that the PSF can be reliably accounted for, even for low $S/N$ galaxies, as residual biases after noise bias correction are only several percentage points.  }
 \label{fig:m_sn_measctr_unres}
\end{figure}

We have also analysed the set of barely resolved images and plotted the multiplicative bias as a function of $S/N$ in Fig. \ref{fig:m_sn_measctr_unres}. These results sketch a similar picture: noise bias can be reduced by roughly half, but not completely removed. However, even with residual noise bias, SNAPG has only percent level biases for very faint, very small objects. This achievement is remarkable and highlights the potential of SNAPG.

\begin{figure}
 \centering
 \includegraphics[width=9cm,height=8cm,keepaspectratio=true]{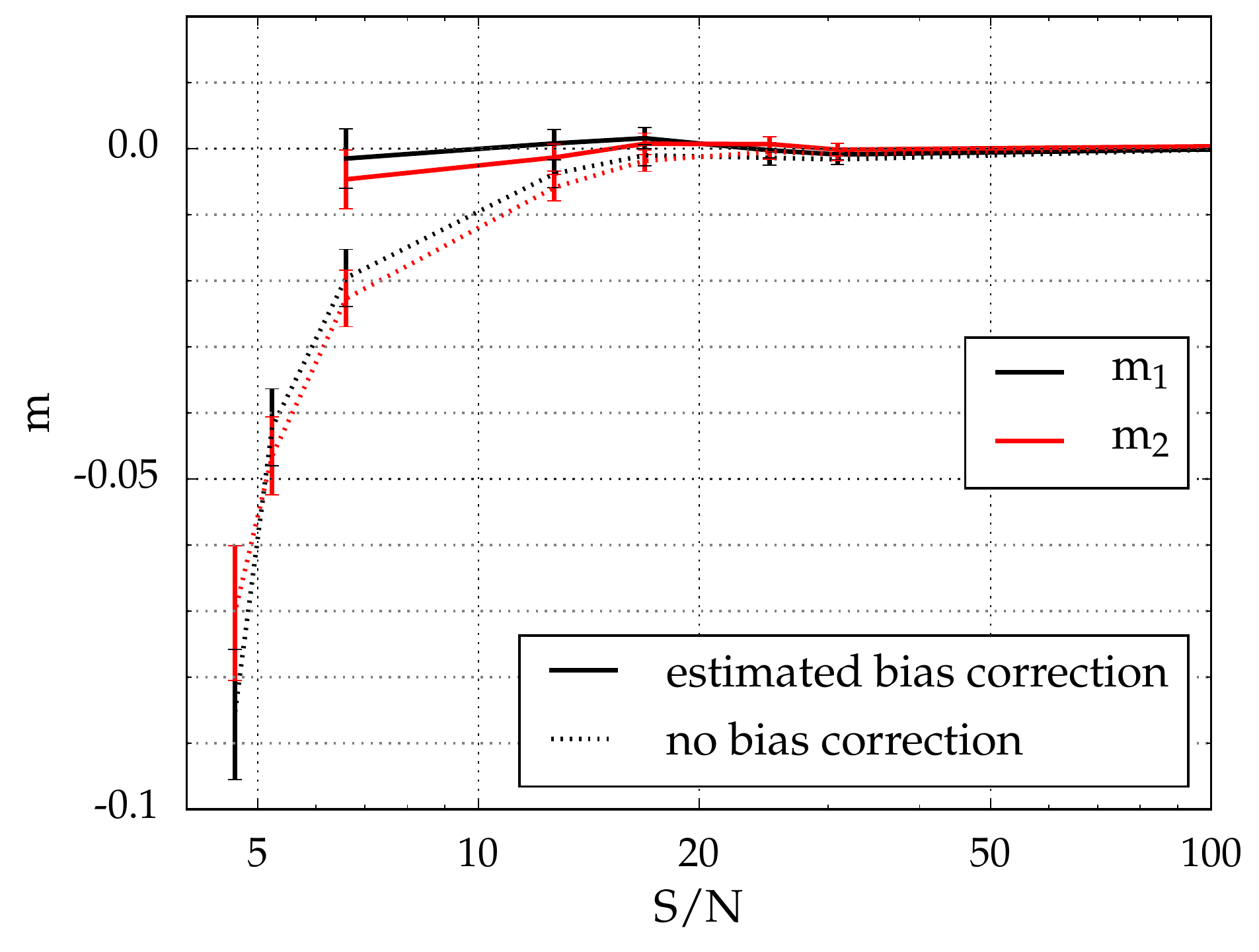}
 \caption{Same as Fig. \ref{fig:m_sn_measctr_res} where the correlation between centroid error and the image noise is removed. Note the different y-axis scale compared to Fig.\ref{fig:m_sn_measctr_res}. The centroid bias correction now accounts for all the bias to subpercent accuracy, until it fails for extremely low $S/N$.}
 \label{fig:m_sn_nocorl_res}
\end{figure}

We have corrected for the non-linearity due to the noisy estimates of the centroid and the second moments themselves have a linear relationship to the noise. Indeed it is shown in Fig. \ref{fig:m_sn_xyvar} that if the centroid bias is perfectly corrected for, the noise in the second moments does not introduce a bias. However, an additional non-linearity in SNAPG remains: the correlation between the centroid error and the pixel noise. This correlation is not present in Fig. \ref{fig:m_sn_xyvar}, but it is in Figs. \ref{fig:m_sn_measctr_res} and \ref{fig:m_sn_measctr_unres}. We now check whether this correlation is the origin of the residual bias after correction, by measuring the centroid and its variance from a different noise realisation than the second moments. We repeat this exercise again for different noise levels and show the results in Fig. \ref{fig:m_sn_nocorl_res}. Even without the use of the correction (dotted lines) the bias is significantly decreased when compared to Fig. \ref{fig:m_sn_measctr_res}, highlighting the bias induced by the correlation. We see that the centroid bias can be corrected to subpercent levels even for galaxies with $S/N <10$. Although the correction breaks down for  galaxies that are too faint, the bias is consistent with zero down to the lowest $S/N$.

\section{GREAT08}
\label{sec:great08}
In the previous section we have shown that the shear estimated using SNAPG is accurate up to the percent level for images with circular Gaussian PSFs. Noise bias can be removed using the centroid correction described in Sec. \ref{sec:bias} if different noise realisations of the same galaxy are available. However, it remains to be shown how SNAPG would perform on real observations and specifically on images without a circular Gaussian PSF. Therefore, we now apply the SNAPG algorithm to the more realistic Gaussianised images of the GREAT08 competition (see Sec. \ref{sec:sims}). These tests will show how well the PSF Gaussianisation algorithm performs. As before we use $a=3$ and $w=3$ for the sizes of the weight functions. For the covariance matrix, $N(\x-\x')$ is given by the auto-correlation function of the kernel which is used to Gaussianise the PSF in the GREAT08 images. This is convolved with the original covariance matrix $\sigma_n^2 \delta (\x -\x')$ and this convolution is used in Eq. \ref{eq:ctrcov} to compute the centroid covariance matrix.

\begin{figure*}
 \centering
 \includegraphics[width=12cm,height=9cm,keepaspectratio=true]{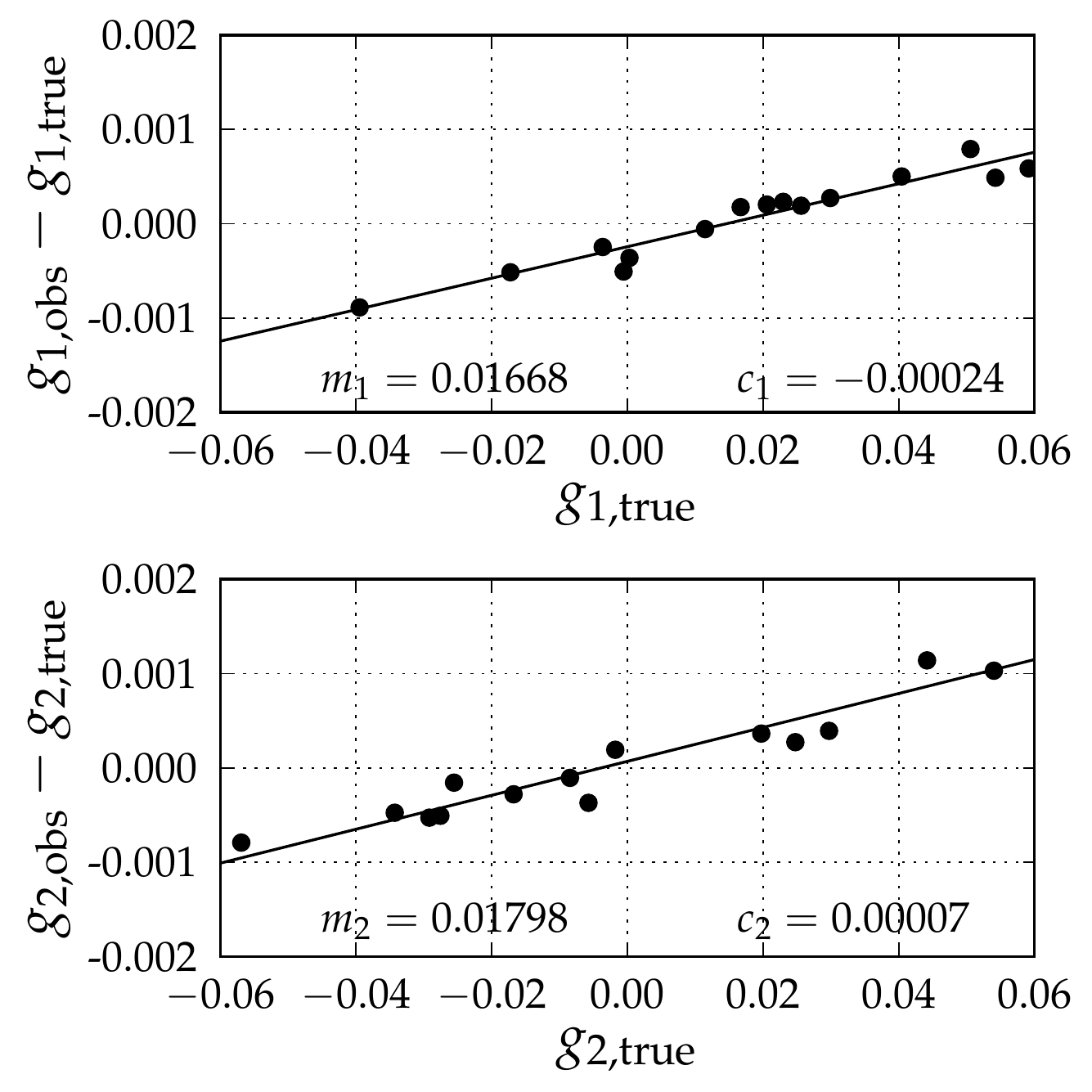}
 \includegraphics[width=12cm,height=9cm,keepaspectratio=true]{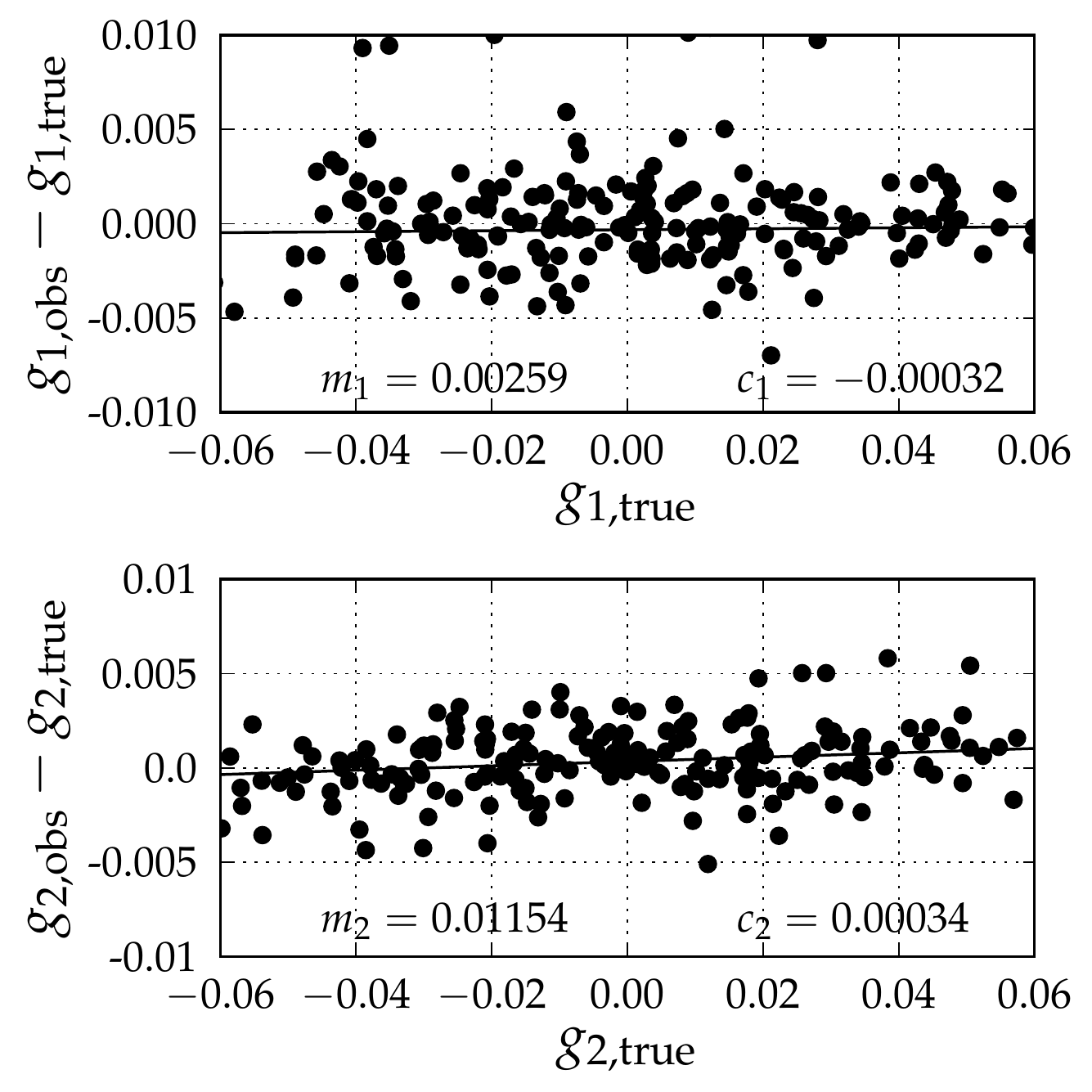}
 \caption{Same set-up as Fig. \ref{fig:high_sn_res}, here for the shear estimates $g_{out}$ for the LowNoise\textunderscore Known images of the GREAT08 challenge outfitted with a circular Gaussian PSF using SNAPG with a correction for correlated noise. The positive multiplicative bias for these high $S/N$ galaxies shows that the PSF Gaussianisation routine did not produce a fully circular Gaussian PSF.}
 \label{fig:glnk_sn}
\end{figure*}

We measure the shear with SNAPG on the 15 Gaussianised LNK images and the results are shown in the left panel of Fig. \ref{fig:glnk_sn}. There is a slight overestimation of the multiplicative bias of 1-2\%, and there is a small additive bias inconsistent with zero $c_1= (2.4\pm 0.4) \hspace{1pt} 10^{-4}$ and $c_2= (0.7\pm 0.5) \hspace{1pt}  10^{-4}$. The sign of the multiplicative bias and the non-zero additive bias point towards a PSF, which is not a circular Gaussian. The results of SNAPG measurements on the 300 RNK images are shown in the right panel of Fig. \ref{fig:glnk_sn}. Again there is positive residual multiplicative bias $m=(+7.5 \pm 2.5) \hspace{1pt} 10^{-3}$, but a slightly lower value than the one we found for the LNK images. This is probably the combination of the Gaussianisation process and the imperfect centroid bias correction we found in the previous section. As we do not possess different noise realisations of the GREAT08 images, we cannot remove the residual noise bias. There is also a small, but statistically significant discrepancy between the bias in $g_1$ and $g_2$ which is not seen in other tests.

We investigate the percent level bias found in the GREAT08 in more detail by looking at the shear bias for various PSF profiles. We simulated two images of galaxies of opposite shears ($g_1=\pm0.03, g_2=\mp0.02$) with a non-Gaussian PSF. Six different Moffat profiles with $\beta=$2, 3, 4, either circular or elliptical, with $\epsilon=+0.02$, $\epsilon=-0.01$ were used as PSFs. The PSF half light radius was 1.76 pixels and the galaxy half light radii were 2.5 pixels, so that these images resembled the \textit{well resolved} images. At $S/N\sim100$ the images underwent PSF Gaussianisation and afterwards the shear was estimated. We found that regardless of the original PSF, the bias in the shear is $\sim$2\%, similar to the results from the GREAT08 images. At such a high $S/N$ this bias is not due to a centroid error and therefore we suspect an imperfect Gaussianisation of the PSF to  be the cause. It is unclear which aspect of the PSF Gaussianisation routine causes the bias in the shear estimate, although it does seem to be robust against variations in the PSF profile.

\section{Discussion}
\label{sec:discussion}

\subsection{SNAPG formalism}
We have introduced the SNAPG formalism and tested its performance as a shear measurement method. For galaxy images convolved with a round Gaussian PSF the effect of shear on weighted second moments of image brightness can be analytically calculated. This analytical treatment is used to create a pipeline which finds the gravitational lensing shear by nulling the polarisations for an ensemble of galaxies. This procedure thus finds an estimate for the shear experienced by the galaxies. On test images with high $S/N$ galaxies convolved with a circular Gaussian PSF, the method obtained shear estimates deviating from the input shears by only parts per thousand.

\subsection{Noise bias}
Like most shape measurement methods, SNAPG suffers a noise bias when applied to images of galaxies with low $S/N$. However, by using only linear combinations of second moments instead of ratios of moments such as the polarisation or ellipticity, much of the noise bias can be avoided. This strategy allows SNAPG to obtain  only a percent level bias in images with a $S/N\approx10$. Noise in the data introduces errors in the centroid estimates, which in turn biases the shear estimates. We compute an analytic treatment to correct the centroids and show that it can significantly improve the performance of SNAPG for low $S/N$ galaxies. Remaining biases after correction for $S/N\approx10$ are in the range of less than one percent.

The residual biases increase with decreasing $S/N$, which indicates that the centroid error correction does not account for the full effect of noise bias. We traced their origin to the correlation between the centroid errors and pixel noise in the second moments. By removing the correlation, we can greatly decrease the measured bias, and also correct for the remaining bias with our centroid bias correction to subpercent accuracy.  For multi-band surveys a possible solution is to use different filters for the estimates of the centroid and the measurement of the moments. In this way, the correlation between the centroid and the image is removed and without this correlation SNAPG can produce almost unbiased results. The impact on the bias of such a scheme will have to be investigated  as galaxy colours and colour gradients may become an issue. In addition, this introduces a correlation with the photometric redshift estimate, which might pose a problem for cosmic shear measurements.

\subsection{Galaxy resolution}
The shape of a galaxy similar in size to the PSF is heavily distorted by the PSF, making it difficult to estimate the intrinsic shape. However, the analytic treatment of the PSF in the SNAPG formalism ensures that shear estimation is possible even for barely resolved galaxies. For galaxies 0.84 times smaller than the PSF, the shear was retrieved to similar accuracy,   as were resolved galaxies for $S/N\approx$7-10 galaxies. By being able to measure unresolved galaxies reliably, SNAPG is able to use the large population of faint small galaxies to boost statistical power.

\subsection{PSF Gaussianisation}
We have run SNAPG on the images of the `LowNoise\textunderscore Known' and `RealNoise\textunderscore Known' branches of the GREAT08 challenge. To make them suitable for SNAPG, the GREAT08 images were first passed through the PSF Gaussianisation. We find a slight overestimation of the shear for the LNK images with $S/N=200$, of the order of 1-2\%. The PSF Gaussianisation introduces a correlation in the noise, which is analytically corrected for. SNAPG can retrieve the shear from the RNK images with $S/N=20$ to an accuracy  that is similar to that for the high $S/N$ images. Further tests revealed that this percent level bias is probably inherent to the PSF Gaussianisation routine that we have used. For a variety of PSF profiles the multiplicative shear bias remained constant around 2\%.
The PSF Gaussianisation appears to be the limiting factor for SNAPG to obtain subpercent shear bias and detailed investigation into this routine is necessary before SNAPG can be reliably applied to observations.

We can compare the performance of SNAPG to the performance of the other methods tested in the GREAT08 challenge. This will only provide an indication as we did not run our pipeline on all datasets in the challenge and shear measurement methods have evolved since. However, a comparison to figures C3 and C4 in \citet{great08} shows that the 1-2\% bias SNAPG has obtained is at least competitive with other shear measurement methods. A more quantitative comparison to other (recent) shape measurement methods will require testing on image simulations which incorporate realistic observational features. However, optimistically the performance we find for SNAPG is sufficient to meet the requirements of the largest cosmic shear survey to date \citep{kids450} without any calibration being required.

\subsection{Shear precision}
\label{sec:precision}
So far we have been concerned only with the accuracy of SNAPG, but an equally valid demand is high precision. To estimate the scatter in the shear estimate we use the simulated images of galaxies observed with the Hubble Space Telescope (HST) included in the GalSim software. These galaxies were observed as part of the COSMOS survey \citep{koekemoer07} and we used galaxies between magnitudes 20 and 24.5, similar to the depths of the Kilo Degree Survey and the Dark Energy Survey. These galaxies were rescaled to a pixel size of 0.214 arcseconds and convolved with a circular Gaussian PSF. We find that the scatter in the shear estimate for this set of galaxies is roughly $0.45/ \sqrt{N_{gal}}$, where $N_{gal}$ is the number of galaxies in the image. Thus the scatter in the SNAPG shear estimate for a fully realistic ensemble of galaxies is worse than an ellipticity based estimate; roughly 2-3 times more galaxies are needed by SNAPG for the same precision. 
This result is more optimistic than the increase by a factor of 10 found by \citet{viola14} in their analysis of a shear estimator based on Stokes parameters. Our use of a weight function reduces the variation of the moments, thereby shrinking the scatter in the Stokes parameter. In our tests we  used identical weights for all sources, which naturally downweighs  large, bright galaxies, which would otherwise dominate the ensemble average of second moments. Ideally, in order to optimise the $S/N$ of the individual moment measurement, the size of the weight function should match the observed size of the galaxy. However, fitting weight functions to individual galaxies is in itself a noisy process that may lead to a bias. We therefore advocate using the same weight function size for all galaxies (since most will be only partially resolved, it is not difficult to find a size that is nearly optimal for most of the galaxies by picking a small multiple of the PSF size; see also Eq. \ref{eq:wfsize}).

A possible improvement is to assign each galaxy a weight to reduce the variance in the shear estimate. We find that for our sample of HST galaxies weighting by the inverse of the true flux can reduce the required number of galaxies by a factor of $\sim$4. This would bring the precision of SNAPG close to the precision of shear estimates based on galaxy ellipticities. In practice, estimating this weight factor from the galaxy fluxes measured in other images (e.g. adjacent photometric bands in a multi-band survey) uncorrelated with the lensing images will avoid introducing noise bias.

\subsection{Variable shear}

Observational weak lensing deals with varying shear fields, for instance in cosmic shear measurements or when measuring the mass of groups or clusters of galaxies. The traditional method is then to average the shear estimate for individual galaxies to obtain the lensing signal. This is not possible with SNAPG as it does not produce a shear estimate per source. In addition, SNAPG requires a large number of galaxies to obtain a precise shear estimate and satisfy the condition that the intrinsic ellipticities average to zero. 

Instead of nulling a single shear value for an ensemble of galaxies, we therefore advocate  nulling a parametrised model shear field for that ensemble. For example, to measure a galaxy-galaxy lensing signal, the model would include parameters that describe the average shear profile of galaxies and their scaling with pertinent galaxy properties. The model parameters would then be varied until the average shear in a number of annular bins around the lensing galaxies is nulled, analogous to a traditional tangential shear stacking analysis. As another example, for cosmic shear measurements, the amplitudes of independent Fourier modes in the shear field could be nulled. \\
Developing this procedure will be left to the future. 

\section{Summary}
\label{sec:summary}
We have presented a new moment-based method that attempts to combine the best aspects of earlier approaches to the problem of high-accuracy, precise shear measurement from galaxy images.
Moment-based methods generally approximate the deconvolution  of the PSF, but do not require any information beyond the data and generally run very fast. Model fitting methods perform exact forward modelling, including convolution with the PSF, but are expensive to run because they need to search through a large parameter space, and may suffer model bias. The shear nulling after PSF Gaussianisation or SNAPG technique deals analytically with the PSF deconvolution and as a moment-based method only requires a few measurements on the data. In addition, SNAPG incorporates a correction scheme to mitigate the effects of noise bias, a major hurdle to all shape measurement techniques.

Idealised test images show that SNAPG can retrieve shears to percent level accuracy for galaxies with low signal-to-noise, even if they are smaller in size than the PSF. The main issue limiting this technique is the correlation between the noisy estimate of the centroid and the pixel noise, which may be mitigated by incorporating further data about the sources, such as images from neighbouring bands in a multi-wavelength survey. In such a set-up, SNAPG can obtain shear estimates to subpercent accuracy for galaxies with a Gaussian PSF.

Application to real data requires PSF Gaussianisation and if this routine is imperfect it can introduce percent level biases. This level of accuracy is comparable to what is required of the shape measurement algorithms used for ongoing surveys. As such, we expect SNAPG to be a useful asset for current and future weak lensing experiments.

\subsection{Acknowledgements}
We would like to thank Massimo Viola, Henk Hoekstra, and Peter Schneider for useful discussions. We also thank the anonymous referee, whose suggestions  helped to improve this paper. RH acknowledges support from the European Research Council FP7 grant number 279396. AB was supported for this research partly through a stipend from the International Max Planck Research School (IMPRS) for Astronomy and Astrophysics at the Universities of Bonn and Cologne and through funding from the Transregional Collaborative Research Centre `The dark Universe' (TR 33) of the DFG. KK acknowledges support from an Alexander von Humboldt Foundation research award.

\bibliography{snapg_lib}

\appendix

\section{Convolution calculations}
\label{app:gconv}

In this Appendix we calculate the result of convolving $[x_ix_jG_1]$ with a Gaussian point spread 
function $G_2$, where $G_1$ and $G_2$ are Gaussians of arbitrary covariance matrix.

First we consider the product of a non-circular Gaussian of covariance matrix $\V$ with an offset one 
of covariance $\pmat$ and centre $\y$:
\begin{equation}
e^{-\frac12 (\x^T\V^{-1}\x)} e^{-\frac12(\x-\y)^T\pmat^{-1}(\x-\y)} .
\end{equation}
The sum $\x^T\V^{-1}\x+(\x-\y)^T\pmat^{-1}(\x-\y)$ can be rearranged to yield
\begin{equation}
(\x-\z)^T\K(\x-\z) - \z^T\K\z + \y^T\pmat^{-1}\y
\label{eq:decomp}
\end{equation}
with
\begin{equation}
\K=\V^{-1}+\pmat^{-1}
\end{equation}
and
\begin{align}
\z = & \K^{-1}\pmat^{-1}\y \\ \nonumber 
   = & (\V^{-1}+\pmat^{-1})^{-1}\pmat^{-1}\y = \V(\V+\pmat)^{-1}\y .
\end{align}
The terms in Eq.~\ref{eq:decomp} not involving $\x$ simplify to
\begin{align}
-\z^T\K\z+\y^T\pmat^{-1}\y = & \\ \nonumber 
                             & -\y^T \pmat^{-1} (\V^{-1}+\pmat^{-1})^{-1} \pmat^{-1} \y \\ \nonumber
                             & + \y^T\pmat^{-1}\y \\ \nonumber
                             & = \y^T (\V+\pmat)^{-1}\y .
\end{align}
Using this result we can calculate the convolution of $x_ix_j e^{-\frac12 \x^T\V^{-1}\x}$ with a 
normalised Gaussian of covariance $\pmat$ as
\begin{align}
\label{eq:xyggconv} 
  & \int d\x\, x_i x_j e^{-\frac12 (\x^T\V^{-1}\x)} {e^{-\frac12(\x-\y)^T\pmat^{-1}(\x-\y)} 
    \over2\pi|\det \pmat|^\frac12} \\ \nonumber
= & \int d\x\, {x_i x_j  \over2\pi|\det \pmat|^\frac12} e^{-\frac12 (\x-\z)^T\K(\x-\z)} 
    e^{-\frac12  \y^T (\V+\pmat)^{-1}\y} \\ \nonumber
= & \left|\det \K^{-1}\over\det \pmat\right|^\frac12 e^{-\frac12 \y^T (\V+\pmat)^{-1}\y} 
    \left(K^{-1}_{ij}+z_iz_j\right)\\ \nonumber
= & \left|\det \V\over\det(\V+\pmat)\right|^\frac12  e^{-\frac12 \y^T (\V+\pmat)^{-1}\y} \times \\ \nonumber
  & \left[\left(\V(\V+\pmat)^{-1}\pmat\right)_{ij} + \left(\V(\V+\pmat)^{-1}\y\right)_i 
    \left(\V(\V+\pmat)^{-1}\y\right)_j\right] \\ \nonumber
= & \left|\det \V \over \det\bmat\right|^\frac12 e^{-\frac12\y^T\bmat^{-1}\y} \\ \nonumber
  & \times \left[\left(\V-\V\bmat^{-1}\V\right)_{ij} +\left(\V\bmat^{-1}\y\right)_i \left(\V\bmat^{-1}\y\right)_j \right]
\end{align}
where in the last line we have defined $\bmat=\V+\pmat$. We note that a deconvolution is simply accomplished 
by changing the sign of $\pmat$.

\end{document}